\documentclass[times,twocolumn,final,authoryear]{elsarticle}
\usepackage{jasr}
\usepackage{framed,multirow}

\usepackage{amssymb}
\usepackage{latexsym}

\usepackage[switch]{lineno}

\usepackage{url}
\usepackage{xcolor}
\definecolor{newcolor}{rgb}{.8,.349,.1}

\usepackage[citebordercolor=white]{hyperref}
\usepackage{aas_macros}
\def\arcsec   {\hbox{$^{\prime\prime}$}}
\def\degr     {\hbox{$^\circ$}}
\def\arcmin   {\hbox{$^{\prime}$}}

\journal{Advances in Space Research}

\begin{document}

\verso{Rajan Chhetri \textit{et al.}}

\begin{frontmatter}

\title{First measurement of interplanetary scintillation with the ASKAP radio telescope: implications for space weather}

\author[1,2]{Rajan \snm{Chhetri}\corref{cor1}}
\cortext[cor1]{Corresponding author: 
	Tel.: +61-8-9266-3577;  
	fax: +61-8-9266-9246;}
\ead{rzn.chhetri@gmail.com}
\author[1]{John \snm{Morgan}}
\ead{john.morgan@curtin.edu.au}
\author[3]{Vanessa \snm{Moss}}
\ead{Vanessa.Moss@csiro.au}
\author[1,3]{Ron \snm{Ekers}}
\ead{Ron.Ekers@csiro.au}
\author[1]{Danica \snm{Scott}}
\ead{danica.scott@postgrad.curtin.edu.au}
\author[3]{Keith \snm{Bannister}}
\ead{Keith.bannister@csiro.au}
\author[4]{Cherie K. \snm{Day}}
\ead{cday@swin.edu.au}
\author[5]{Adam T. \snm{Deller}}
\ead{adeller@astro.swin.edu.au}
\author[5]{Ryan M. \snm{Shannon}} 
\ead{rshannon@swin.edu.au}

\address[1]{International Centre for Radio Astronomy Research, Curtin University, GPO Box U1987, Perth, WA 6845, Australia}
\address[2]{CSIRO Space and Astronomy, P.O. Box 1130, Bentley, WA 6102, Australia}
\address[3]{CSIRO Space and Astronomy, P.O. Box 76, Epping, NSW 1710, Australia}
\address[4]{Department of Physics, McGill University, Montreal, Quebec H3A 2T8, Canada}
\address[5]{Centre for Astrophysics and Supercomputing, Swinburne University of Technology, John St, Hawthorn, VIC 3122, Australia}
\received{6 Apr 2022}
\finalform{6 Apr 2022}
\accepted{xx XXX 20XX}
\availableonline{XX Xxx 20XX}
\communicated{S. Sarkar}

\begin{abstract}
%%%
We report on a measurement of interplanetary scintillation (IPS) using the Australian Square Kilometre Array Pathfinder (ASKAP) radio telescope. Although this proof-of-concept observation utilised just 3 seconds of data on a single source, this is nonetheless a significant result, since the exceptional wide field of view of ASKAP, and this validation of its ability to observe within 10 degrees of the Sun, mean that ASKAP has the potential to observe an interplanetary coronal mass ejection (CME) after it has expanded beyond the field of view of white light coronagraphs, but long before it has reached the Earth.
We describe our proof of concept observation and extrapolate from the measured noise parameters to determine what information could be gleaned from a longer observation using the full field of view.
We demonstrate that, by adopting a `Target Of Opportunity' (TOO) approach, where the telescope is triggered by the detection of a CME in white-light coronagraphs, the majority of interplanetary CMEs could be observed by ASKAP while in an elongation range $<$30\degr.
It is therefore highly complementary to the colocated Murchison Widefield Array, a lower-frequency instrument which is better suited to observing at elongations $>$20\degr.
%%%%

\end{abstract}

%\textsl{}

\end{frontmatter}

%% main text
\section{Introduction}
Interplanetary Scintillation (IPS) was discovered by \citet{Clarke:phdthesis}, who postulated that the phenomenon may be associated with the Solar Corona. 
The initial use of IPS by \citet{1964Natur.203.1214H} was as a technique for identifying compact ($\lesssim$0.3\arcsec) astrophysical radio sources.
Later it was confirmed that enhancements in the observed scintillation index were associated with solar flares \citep{1967Natur.213..377S}, and over the following decades,  IPS observations played a vital role in uncovering the nature of the solar wind, including measurement of the solar wind velocity beyond the plane of the ecliptic \citep{1967Natur.213..343D}, observation of the acceleration of the solar wind close to the Sun \citep{1971A&A....10..310E}, and changes in the solar wind over the solar cycle \citep{1980Natur.286..239C}.

More relevant to this work, \citet{1968PASA....1..142D} used a ``grid'' of IPS sources to reconstruct the path of an enhancement in the solar wind, using both scintillation indices and power spectra.
Later, \citet{1982Natur.296..633G} (see also \citealp{Vlasov1979}) proposed that a network of IPS sources be monitored daily, so that the changes in scintillation index could be used to track interplanetary disturbances as they move outwards through the Heliosphere.
These findings motivated the construction of dedicated IPS arrays \citep[e.g.][]{2011RaSc...46.0F02T}, whose data can be used in near-real time to reconstruct the inner heliosphere. \citep{1998JGR...10312049J,2013PJAB...89...67T}.

In addition to purpose-built instruments, most radio telescopes can be used to make IPS measurements, the main requirement being sub-second time resolution.
We have shown that the Murchison Widefield Array \citep[MWA;][]{2013PASA...30....7T} is an outstanding instrument for IPS studies \citep{2018MNRAS.473.2965M} due to its extremely wide field of view ($\sim$900 sq. deg. at 162\,MHz) and its ability to make high-fidelity images at high time resolution.
In just a few minutes, we can make simultaneous IPS observations across hundreds of sources and several frequency bands (the MWA has 30.72\,MHz of instantaneous bandwidth which can be deployed flexibly across the observing frequency range of 75--300\,MHz).

Also located at the Murchison Radio Observatory (MRO) in Western Australia, the Australian Square Kilometre Array Pathfinder radio telescope (ASKAP) is an array of 36 $\times$ 12\,m dishes operated as part of the Australia Telescope National Facility. Each of the 36 dishes is equipped with a 188-element phased array feed (PAF) receiver, which widens the 36-beam field of view to approximately 5 $\times$ 5 degrees and enables ASKAP to be an excellent wide-field, high-speed survey instrument. The frequency range of ASKAP is 700--1800\,MHz, with a current instantaneous bandwidth of 288\,MHz, a standard channel resolution of 18.5\,kHz and a standard integration time of 10\,s. The ASKAP telescope and encompassing system elements are fully described in \cite{2021PASA...38....9H}. 

Since ASKAP shares the key characteristics of wide field of view and high fidelity imaging capability with the MWA, we were keen to assess the potential of ASKAP for making IPS observations using the same widefield imaging approach that we have pioneered with the MWA.
ASKAP observes at higher frequencies than the MWA, and so is more suited to making observations closer to the Sun (5\degr--30\degr or so), making it complementary to the MWA, which is better suited to observing beyond 20\degr\ elongation.
Since ASKAP's standard 10\,s correlator integration time precludes IPS observations, we used the alternative pathway offered by the Commensal Real-time ASKAP Fast Transients (CRAFT) System \citep{2010PASA...27..272M}.
This system has the advantage that it offers extremely high time resolution, but the disadvantage that (for now at least) only a very short time interval can be captured.
Notwithstanding this limitation, we were able to use this system to unambiguously detect IPS, demonstrating both that the presence of the Sun (an extremely bright $\sim 10^6$\,K source at ASKAP frequencies) does not present any obvious problems, and that the instrument is sufficiently stable on the relevant timescales for reliable IPS measurements to be made.
Below, we describe this detection in detail as well as exploring the potential for ASKAP as a space weather monitoring facility.
The paper is presented as follows: in Section 2 we present the details of our ASKAP observation and results. In section 3, we present our analyses of the results. 
Finally, in Section 4 we describe a pathway to regular IPS observations with ASKAP and it's implications.

\section{Observations and Results}
\subsection{Observations}
\label{Sec:Observations}
The CRAFT system was developed to detect fast ($<$5\,s) transient radio sources \citep{2010PASA...27..272M}, and is primarily used to detect and localise Fast Radio Bursts \citep{Macquart2020Natur.581..391M, Heintz2020}. Upon receipt of a trigger, the system downloads voltage data from a specified beam of each individual antenna to correlate offline \citep{Bannister2019}. Since the subsequent processing takes place offline using the Distributed FX (DiFX) software correlator \citep{DIFX}, arbitrarily short integration times are possible. However, the limited memory available for buffering voltage data, combined with the high data rate, means that the voltage download duration is limited to a maximum of 3.1\,s when the data are stored at the standard 4-bit precision.

With respect to conducting IPS observations using ASKAP, there are a few considerations in terms of how close to the Sun ASKAP can point, how quickly it can get on source, and how science operations are generally conducted with the telescope. When pointing at the Sun in rainy conditions, the wet surface of the parabolic ASKAP antenna becomes reflective rather than diffusive to optical wavelengths, and focused solar radiation can damage the PAF. Due to the consequent risk associated with solar observations, ASKAP science observations are limited to field centres beyond a solar elongation of 10 degrees. However, since the telescope has a wide field of view, such a pointing will cover a 5 degree range of elongations centred on 10 degrees. Each ASKAP dish, due to its relatively small size, can slew rapidly to position, reaching any pointing within a few minutes. 

In order to select a candidate source for our test observation, we examined strongly scintillating sources from our MWA IPS catalogue (Morgan et al. in prep) which were expected to be bright at ASKAP frequencies as indicated by the 1.4\,GHz flux density listed in the NRAO VLA Sky Survey \citep[NVSS;][]{1998AJ....115.1693C}. NVSS 070029+190541, our chosen candidate source, has a flux density in NVSS of $\sim$400mJy and at MWA frequencies has an IPS scintillation index consistent with an unresolved source.
NVSS 070029+190541 was observed on 25 June 2021 at a solar elongation of 11.2 degrees. For this proof-of-concept observation, only the data from the PAF beam with our target source in it was preserved.

The methods of correlation, calibration, and imaging were very similar to those described in \cite{Bannister2019} and \cite{2020MNRAS.497.3335D}. Voltages were downloaded for both linear polarisations for a single PAF beam across 24 antennas.
The voltages were then correlated with an integration time of 100\,ms, 
yielding a total of 31 timesteps over 3.1\,s duration. Antenna-based, frequency-dependent phase and flux calibration solutions were derived from a similar set of voltages downloaded during an observation of PKS 0407$-$638 made on the same day as our target source.

\subsection{Results}
\begin{figure}%{r}{0.\linewidth}
	\includegraphics[width=0.48\textwidth]{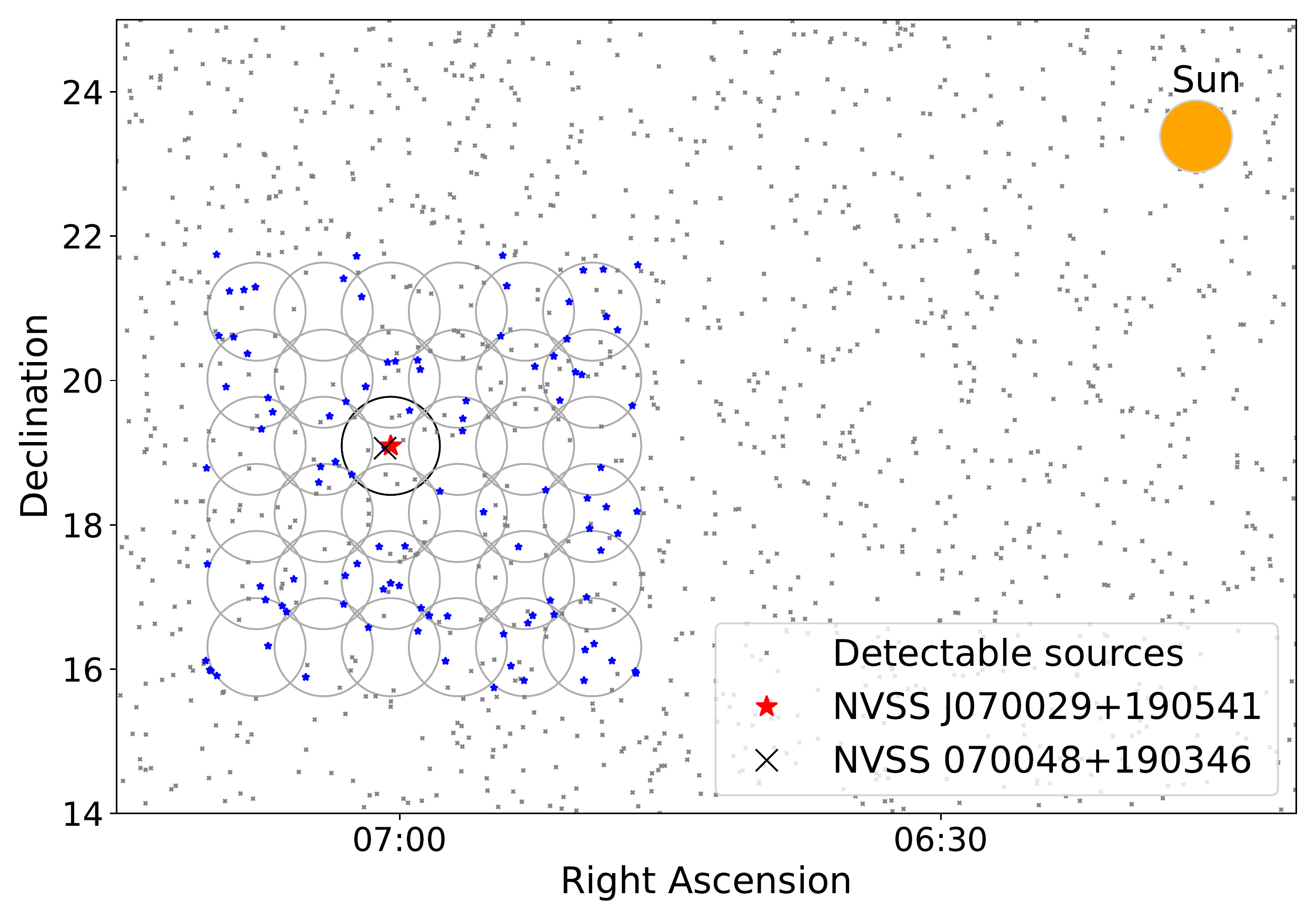}
	\caption{{Schematic of our IPS detection in the ASKAP field of view from 36 PAFs shown as overlapping circles. Only one PAF data was downloaded, which is indicated with the dark circle. The position of our target source and the second detected source are indicated with the red star and a black cross respectively. Sources that could be detectable above 5$\sigma$ in 200\,ms images, based on RACS survey, are shown with grey dots. To demonstrate the predicted density of IPS sources in 1 minute of ASKAP observations, based on counts of flat-spectrum sources, we randomly highlight 131 RACS sources using blue stars.}}
	\label{Fig:ASKAPfov_Sun}
\end{figure}

We then made individual Stokes-I images for each of the 31 timesteps correlated with angular resolution of 25.0\,arcsec (major axis of synthesized beam).
We also detected another source (NVSS J070048+190346) at a separation of 4.8\arcmin\ ($\sim$11.5 resolution elements) from the target source with an average S/N of 4.6. The schematic of the sky coverage with respect to the position of the Sun for our observation is shown in Figure \ref{Fig:ASKAPfov_Sun} . The non-target source is not a known scintillator at 162\,MHz, and it is not sufficiently bright for our current MWA IPS survey data to provide a strong constraint on any scintillation.
These two sources were the only two clearly visible in our images.

Visual inspection of the images confirms that they are not affected by Radio Frequency Interference (RFI) or solar radio bursts.
The point spread function of the bright target source is stable, and there are no other artefacts.
We conclude from this that there are no obvious instrumental errors,
which is as expected, since the CRAFT system is a well-established and such errors would compromise its scientific productivity.

\begin{figure}%{r}{0.\linewidth}
	\includegraphics[width=0.48\textwidth]{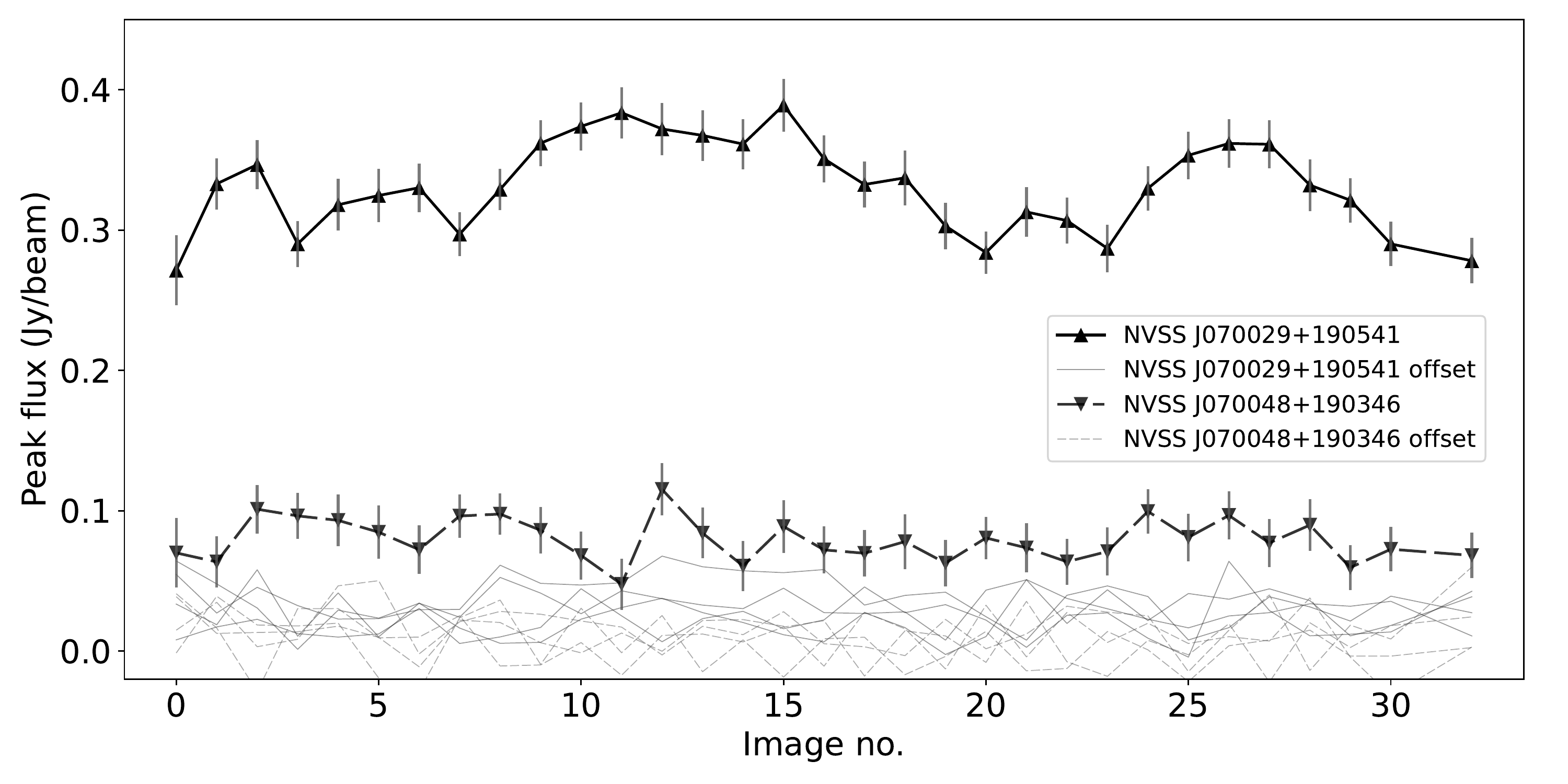}
	\caption{{Time series for two sources detected in our images. Known scintillator NVSS 070029+190541 is the stronger source of the two, shown with upward facing triangles connected with solid lines. Source NVSS 070048+190346 is downward facing triangles connected with dashed lines. Error bars give the image RMS at the source's location in the image. Time series obtained for four pixels per source, offset by 100 pixels in different directions, are shown with faint lines.}} 
	\label{Fig:time_series}
\end{figure}

We measured the flux densities and peak brightness for the two objects by fitting an elliptical Gaussian (major axis: 25$\arcsec$.02, minor axis: 16$\arcsec$.64, pa: -17.7$\degr$) to their positions in each 98\,ms image. The time series of brightness thus produced is plotted in Figure \ref{Fig:time_series}. The uncertainties in brightness are the image RMS (estimated by taking RMS of the pixel values in a large part of each image, away from any radio source). The plot also shows time series for four offset pixels (100 pixels away from source positions) in four directions for each of the two objects, which represent brightness values purely due to noise in the image. Since there is no noticeable correlation in the time series of the two sources or between the sources and the respective offset pixels, we can be confident that the variation in brightness is not due to calibration issues. 

\begin{figure}%{r}{0.\linewidth}
	\includegraphics[width=0.48\textwidth]{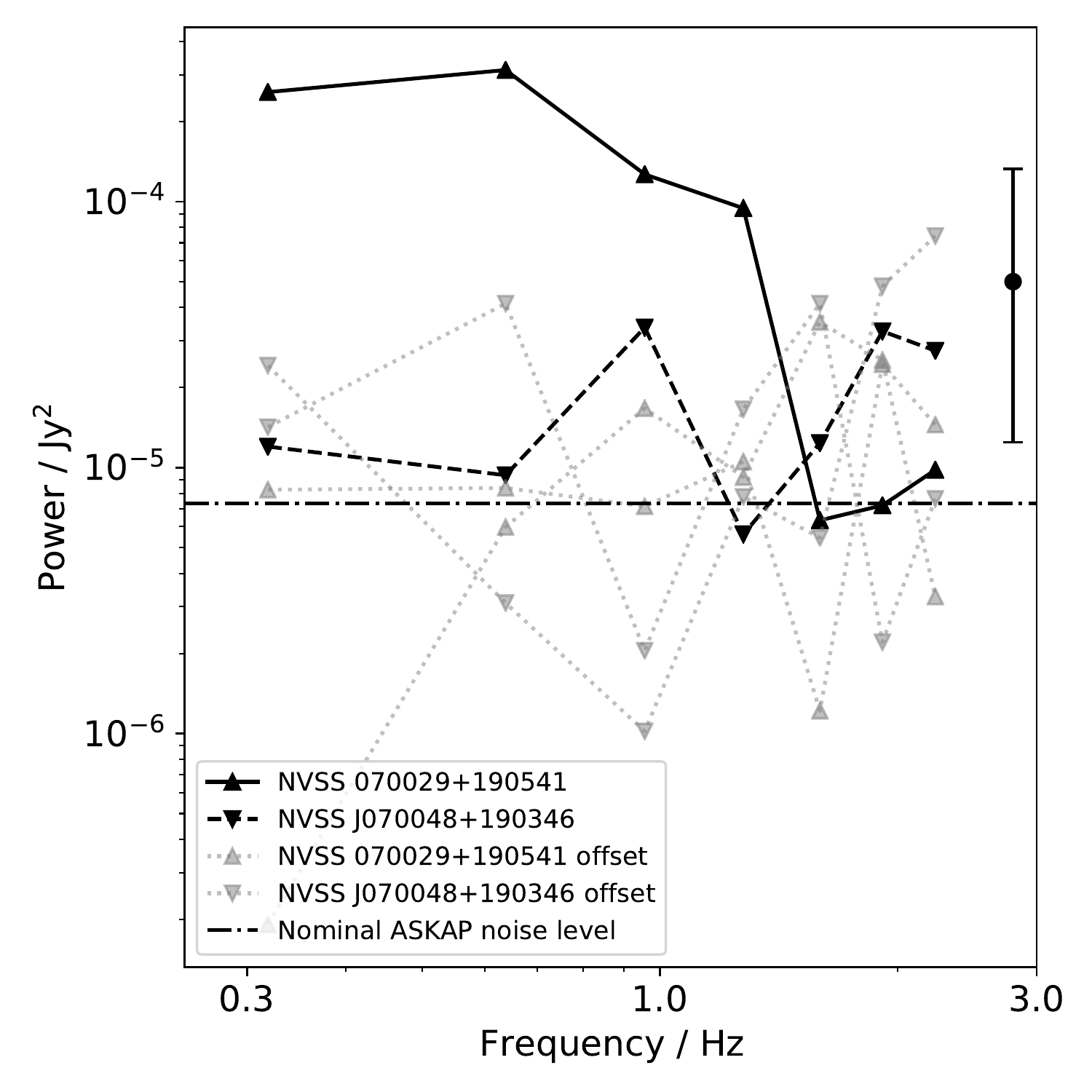}
	
	\caption{{Power spectra for both the scintillating and non-scintillating sources. Adjacent pairs of time series values have been averaged to focus on the region of the spectrum of interest, and the DC and Nyquist channels are not plotted. The error bar shows the 68\% (i.e. approx 1-sigma) confidence interval about the median for the 2 degree-of-freedom $\chi^2$ error distribution which results for a power spectrum derived from a single Discrete Fourier Transform \citep{welch:1967}. However, note that this distribution is not centrally peaked and has a mode of zero.}} 
	\label{Fig:power_spec}
\end{figure}

The power spectra of the time series plotted in Figure~\ref{Fig:time_series} are plotted in Figure~\ref{Fig:power_spec}. While the errors are necessarily large given the very short time series, there is no suggestion of excess on-source variability for the non-IPS source, so instrumental effects, such as unstable gain are unlikely to be responsible for the excess variance observed for the IPS source. Moreover, the power spectrum for the IPS source matches that expected for scintillation in the weak regime, with the `Fresnel Knee' located just below 1\,Hz. The location of this knee scales with the Fresnel scale, which scales with wavelength, $\lambda$, as $\sqrt{\lambda}$ \citep{Narayan1992}, and indeed this power spectrum appears shifted higher in frequency by a factor of $\sim$2 compared to those observed at MWA frequencies \citep[see][Figure~1]{2018MNRAS.473.2965M}. 

We note that the IPS signature in this case is fully resolved with 200\,ms time resolution, so we have averaged to this resolution to generate the power spectrum (the power spectra without this averaging step are all consistent with white noise above 2.5\,Hz).
It is conceivable that higher resolution may be required to capture the IPS power spectrum for very high solar wind speeds (since this will shift the IPS signal to higher frequencies).

\citet{McConnell2020} report the sensitivity of ASKAP as a function of observing frequency (Figure 1). We use this to estimate an expected image RMS (due to system noise) of 5.13 mJy/beam for images made at 200-ms cadence.
A line representing white noise at this level is also plotted on Figure~\ref{Fig:power_spec} and is consistent with the noise level that we observe (both on the non-IPS source, off source, and at high frequencies on the IPS source where the IPS signature is fully resolved).
 
From the time series of our target source, we estimate its scintillation index to be 0.098. \cite{Rickett1973} provides the empirical relationship $m_{pt} = 0.06\lambda^{1.0}p^{-1.6}$ which gives the expected scintillation index ($m_{pt}$) as a function of wavelength ($\lambda$) and point of closest approach of line of sight to the Sun in au (p) for a point source. At our central observing frequency of 863.5\,MHz, the maximum scintillation index expected for a point-like object is 0.28 . This indicates that the source scintillation is slightly resolved at these frequencies (in contrast to the unresolved scintillation index observed at MWA frequencies). This is perhaps expected since Very Long Baseline Interferometry (VLBI) observations of peaked spectrum sources show a double morphology, and the spatial scales probed by IPS are smaller at higher frequencies. As well as hinting at the potential astrophysical utility of IPS observations with ASKAP, this finding also underscores that the IPS characteristics of sources can change at different frequencies, and so care should be taken when translating IPS catalogues from one frequency to another.

Nonetheless, our central finding is that we are able to report the first detection of interplanetary scintillation with the ASKAP telescope, demonstrating the potential to use ASKAP to probe the solar wind to within $\sim$ 42 solar radii.

\section{Analysis}
As noted above, our test image was constructed from only one PAF beam out of 36, and the total length of the observation was 3.1 seconds.
In the future, using a new system in development (see Section.~\ref{sec:operations}), we expect to be able to use all 36 PAFs to obtain a wide field of view ($\sim$6$\times$6 sq. deg.) to make IPS observations of much longer duration.
If we centre the FoV at the solar elongation of 11.5 degrees, as we did with our test observation, we expect a typical solar elongation coverage between 8.5 and 14.5 degrees per pointing. 

\subsection{Potential for ASKAP to perform Widefield IPS measurements}
In \cite{2019PASA...36....2M}, we showed that noise in the `variability image' (i.e., the image made by taking the standard deviation of each pixel from each image in the time series) increases as $t^{1/4}$. For a sensitivity in a single 200\,ms image of 5.13\,mJy, we expect a single minute long observation to have an 5$\sigma$ detection limit of 4.12\,mJy. Since flat-spectrum sources are the dominant compact objects at gigahertz frequencies and above, we used the 5\,GHz counts of flat-spectrum objects \citep{Condon1984a} to estimate that approximately 131 IPS sources would be detected in a single 1-minute observation with ASKAP, equating to 5.2 sources per square degree. 
This is an extremely high density of sources, far exceeding any IPS observations taken up to now.

In MWA observations as yet unpublished (Morgan et al. in prep.), we have detected CMEs in our wide-field IPS data. 
They typically present as an enhancement of the ``g-level'' \citep[the ratio of the observed scintillation index to baseline,][]{1982Natur.296..633G} over an arc roughly equidistant from the Sun, a few degrees in thickness.
With 131 sources detected across the field of view in a single pointing, it should be possible to localise solar elongation of the CME to degree-level accuracy.
10s of sources would be expected to lie within the CME, and power spectra could be generated for each, from which solar wind parameters such as the velocity \citep{1990MNRAS.244..691M}, and the power law index of the turbulence \citep{1983A&A...123..207S} can be determined, at least for the stronger detections. For example, \citet[][]{2015SoPh..290.2539M} have estimated solar wind speed in the inner heliosphere by model fitting to power spectra of sources observed with high S/N. We estimate 2.5 sources per square degree with S/N$\geq$10 in the ASKAP field of view which can be used to for such studies.

The high density of sources probed simultaneously also enables an alternative approach to velocity determination: with observations of the same field spaced a few hours apart, motion of the CME (typically $\sim1^\circ$ per hour) across the sky should be directly discernible.

\subsection{Determining the baseline scintillation level of ASKAP IPS sources}
In order to make measurements of the g-level of an IPS source, it is necessary to have a measurement of the source's baseline scintillation level.
This will be different for each source based on its sub-arcsecond structure.
Eventually we hope to perform an IPS survey of the entire ecliptic using ASKAP, which would provide the required information for each source.
However, until we have such a survey there are a number of strategies that could be employed to extract Space Weather information from ASKAP IPS observations.
First, we can rely on other IPS observatories (including the MWA) to provide us with lists of known IPS sources and their properties.
However, this negates the most unique feature of ASKAP observations: the very high number of sources ASKAP can detect (thanks to its high sensitivity and our widefield approach); unfortunately this means that most of these sources will not be known IPS sources.

The flat spectrum of an extragalactic source is a reliable indicator of its very compact nature \citep[e.g.][]{Petrov2019, moldon_2015A&A...574A..73M, Jackson_2016A&A...595A..86J, Jackson2022A&A...658A...2J}, especially at high radio frequencies where the extended steep spectrum components become weak. Our study of flat-spectrum sources at 200\,MHz (Chhetri et al. under review with MNRAS) using the GLEAM catalogue finds that 6.3\% of overall sources show flat-spectrum. We can use this number to estimate that $>$20\% of total sources at ASKAP frequencies will be flat-spectrum sources, which is in line with findings of other studies \citep[e.g.][]{Condon2009}. This distribution of flat-spectrum, hence, compact sources further supplemented with other approaches such as MWA IPS measurements (Morgan et al. in prep.) and existing VLBI catalogues \citep[e.g.][]{Petrov2019} will provide a dense network of IPS sources at ASKAP frequencies.

Finally, when observing a CME event, ASKAP would make multiple observations spaced $\sim$1 hour apart, and for each pair of observations adjacent in time, the ratio in scintillation index can be determined for each source.
This approach is similar to the ``running difference images'' which are so often employed in analysis of white light coronagraphs \citep[e.g.][]{2004A&A...425.1097R}
The additional advantage in this case, is that in the ratio of g-levels, the baseline level of scintillation cancels out, and so construction of these `g-ratio' maps do not require a reference catalogue.

\subsection{Observing Space Weather with ASKAP: a `target of opportunity' approach}
\label{sec:too}
IPS can be used to observe the ambient background solar wind, as well as for observing a range of solar wind phenomena including interplanetary CMEs, as well as interactions between the fast and slow solar wind \citep[e.g.][]{1997PCE....22..387B}.
There are also a number of potential strategies for detecting and characterising significant space weather events with ASKAP, including blind searches.
Here, we focus on one potential approach for using a modest amount of ASKAP observing time to provide information that may be a useful input to Space Weather forecasts: using ASKAP to follow up the detection of a CME in Large Angle Spectral Coronagraph images \citep[LASCO;][]{1995SoPh..162..357B} using the Computer Aided CME Tracking algorithm \citep[CACTus;][]{2004A&A...425.1097R,2009ApJ...691.1222R}.
We assume this approach for our case study here, since 
1) ASKAP is a multi-user instrument, and so using other instruments dedicated to solar observations to ensure maximum utility of any ASKAP observations is appropriate;
2) CACTus and its accompanying catalogue is widely used, and timely alerts of significant CMEs are broadcast to the Space Weather community; 
3) we have used exactly this technique to detect CMEs in MWA IPS data (Morgan et al. in prep.).
Alternative approaches are discussed briefly in Section~\ref{sec:discussion}.

Essentially, CACTus provides, among other parameters, the launch time of the CME and its radial velocity away from the Sun in the plane of the sky in km/s.
The latter can readily be converted back to angular speed (in degrees per hour).
For our purposes, since ASKAP has such a wide field of view, we can assume that this angular speed will be constant as the CME moves away from the Sun (i.e., we neglect complications such as the 3D motion of the CME and changes in speed, e.g. due to the ambient wind), though in principle we could adopt a more nuanced model.
Our proof-of-concept observation has demonstrated that we can make IPS observations at 11.2\,degrees elongation, and there is no reason to think that we cannot make IPS observations from 7.5\degr\ from the Sun, where IPS enters the weak regime  all the way out to 30 degrees, by which point the scintillation index drops to 0.06.

\begin{figure}
	\includegraphics[width=0.48\textwidth]{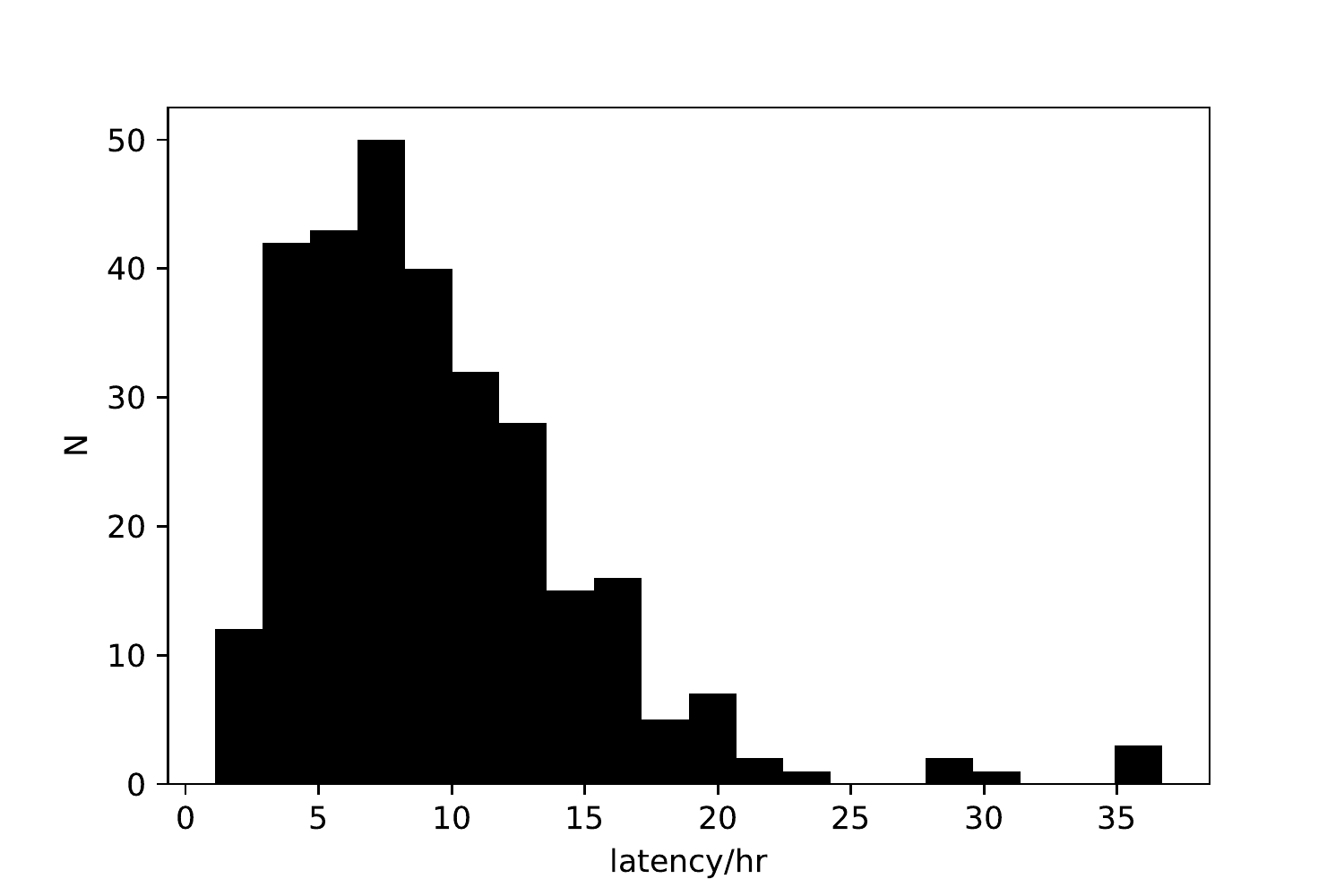}
	\caption{Histogram of latencies of the CACTus system. 50th, 75th and 90th percentiles are 8.3\,hr, 12\,hr and 16.3\,hr, respectively.} 
	\label{fig:latency}
\end{figure}

We wish to determine what fraction of CACTus-detected CMEs would, in principle, be observable with ASKAP.
This depends on the typical latency between CME launch time and the CACTus alert being disseminated in addition to whether or not the Sun (and the CME) is above the horizon at this time, or shortly after.
In order to assess the latency of the LASCO/CACTus alert system, we monitored the status of the CME alert webpage continuously for a 6-month period (2021-Sep--2022-Mar).
During this time, a new alert was issued 299 times.
On each of these updates, we recorded the time that the alert was issued and the estimate of the CME launch time.
The difference between these two times is the latency, the distribution of which is shown in Figure~\ref{fig:latency}.

From the velocity and launch time provided by CACTus, we can determine, for each CME, the time at which it will reach each elongation in the range 0--30\,degrees.
\begin{figure}
	\includegraphics[width=0.5\textwidth]{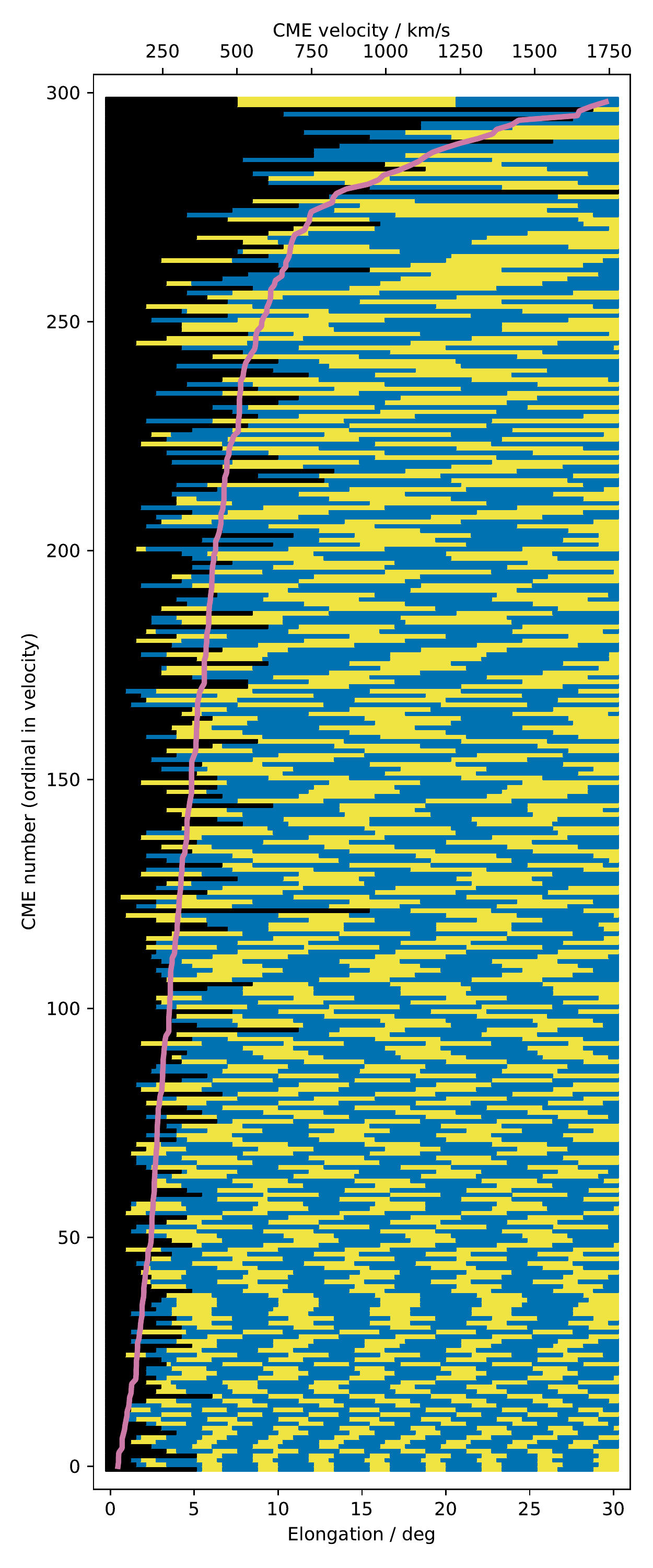}
	\caption{Lower x-axis: visibility of each of 299 CMEs at ASKAP as they propagate from 0--30 deg solar elongation. Black indicates that an alert has not yet been issued for CME detection. Blue indicates the Sun is below 10\degr\ elevation at the location of ASKAP. Yellow indicates that the Sun is above 10\degr\ elevation. Upper x-axis: CME initial velocity (pink line).} 
	\label{fig:cme_prop}
\end{figure}
This is shown in Figure~\ref{fig:cme_prop}, where each time for each CME as it propagates from 0--30\,degrees, it is indicated whether the Sun is up at the ASKAP site.
Almost all CMEs would be observable with ASKAP (many on consecutive days), the only exception being the fastest-moving (velocity, $v>$1000\,km/s) CMEs with higher latency, which are either beyond 30\,degrees before the alert is even issued, or the alert is issued during the night (for ASKAP), with the CME moving beyond 30\,degrees by dawn.
Using the distribution of latencies in Figure~\ref{fig:latency}, it is possible to be a little more quantitative about what fraction of CMEs would be observable (it is also useful to know that CACTus alerts are currently issued every 3 hours starting at 01:30 UTC).
For all but the 10:30, 13:30, and 17:30 (all UTC) alerts, the Sun is up, or will be shortly.
Thus 5/8 of CMEs will be observable for all unless they are extremely fast ($v>$1500\,km/s) and/or high latency (above 90th percentile).
For the 10:30 UTC alert, CMEs will be observable unless they are fast ($v>$1000\,km/s) and/or at median latency.
For the 13:30 UTC alert, CMEs will be observable unless they are fast ($v>$1000\,km/s) and/or moderately high latency (above 75th percentile).
For the 16:30 UTC alert, CMEs will be observable unless they are fast ($v>$1000\,km/s) and/or very high latency (above 90th percentile).

We should note that the fastest events that are the most damaging at Earth, and these can take less than one day to the 1 AU distance to Earth.
A large proportion of the small number of CMEs that are not observable with ASKAP could easily be observed by the MWA, which is colocated with ASKAP, and can make IPS observations in the range 20--50~degrees.
This is discussed further in section~\ref{sec:discussion}.
However we reiterate that the very fastest CMEs may not be observable by ASKAP or the MWA, unless very fortuitously timed, and this provides strong motivation for multiple IPS observatories located at different longitudes.
\section{Discussion}
\label{sec:discussion}
\subsection{Towards Operational IPS observations with ASKAP}
\label{sec:operations}
The mode used for the observation presented here serves for demonstrating the capability of ASKAP for IPS observations but is not suited to operational use, primarily because it only allows such a short observing time.
This is because the mode was developed for capturing Fast Radio Bursts, which are very short duration (compared to an IPS observation).
On the other hand, IPS observations do not require the same time or spectral resolution;
the 100\,ms duration used here is more than adequate (in fact we used 200\,ms to generate the power spectrum in Figure~\ref{Fig:power_spec}).
This raises the possibility of developing a new system which would allow us to record ASKAP visibility data with sufficient frequency resolution for imaging \citep{Perley1981, 2021PASA...38....9H},
sufficient time resolution $\sim100$\,ms for IPS, and a sufficiently long observing time (up to a few minutes).

Fortunately, such a system is currently being developed as an upgrade to the ASKAP FRB detection system that will process high time resolution visibility data. This system, known as CRACO (the CRAFT Coherent system), will be able to record visibility data products at the required 100\,ms time and 1\,MHz frequency resolution to disk. The data can be recorded continuously, up to the limits of disk space, which should be several hours. Once recorded, these data can be calibrated, imaged, and processed offline, in roughly the same manner as detailed above.

ASKAP is currently in the process of transitioning from commissioning to full survey operations, the latter of which are expected to start towards the end of 2022. Science operations for ASKAP are focused on maximising the automation and autonomy of the telescope while minimising the reliance on human decisions at any point in the system and ensuring high scientific data quality (Moss et al. in prep). This approach extends also to the scientific scheduling of the telescope, which is primarily managed by SAURON (Scheduling Autonomously Under Reactive Observational Needs) with manual input. SAURON is designed to make decisions autonomously and dynamically based on the pool of pending observations, the current state of the system, the status of the surrounding environment, and any associated weightings or priorities that feed into the decision-making process. In the context of scheduling IPS, this means that long-term we expect to be able to automatically ingest triggers and incorporate them into scientific scheduling with minimal human oversight and minimal observational disruption.

\subsection{The potential of ASKAP IPS measurements to contribute to Space Weather research and forecasting}
As demonstrated here, ASKAP has the potential to track interplanetary CMEs once they have left the field of view of most heliographs, but long before the CME is approaching the Earth.
Being ground-based, ASKAP is limited to daytime observing of IPS. However, as we have shown in Section~\ref{sec:too}, the vast majority of CMEs will still be observable.
The small number of CMEs that will be missed (due to extreme speed or slow alerts) can be observed by the MWA, a similarly wide field of view instrument colocated with ASKAP.
Since the MWA operates at a lower observing frequency of 70--300\,MHz, it is better suited to observations further from the Sun.
Furthermore, radio observations are much less impacted by weather events than observations at other wavelengths (such as optical observations), and even quite severe ionospheric conditions have limited effect on IPS observations, especially at ASKAP frequencies.

We anticipate that ASKAP would be used occasionally to track significant space weather events, leaving routine monitoring of the Sun and solar wind to dedicated instruments.
We expect that further development of the TOO capability of ASKAP via SAURON should enable IPS observations to be carried out in a prompt and automated way.

Any number of observations or models could be used to trigger ASKAP IPS observations.
In section~\ref{sec:too}, we discuss one possibility in detail.
For ground-based triggers, geographical considerations might influence the choice. 
For example, the K-Cor white-light Coronagraph \citep{2017SpWea..15.1288T} is 6 hours ahead of the Murchison Radio Observatory in longitude, so a rapidly-moving CME discovered very close to the Sun could be followed up as it moves further away from the Sun at a later time.
Similarly, the ISEE telescope, a dedicated and very well-established multi-station IPS telescope in Japan \citep{2011RaSc...46.0F02T}, is just 1 hour ahead of ASKAP and has already been used to provide reference for MWA IPS studies \citep{2018MNRAS.473.2965M}.
ISEE provides rapid information on IPS sources (with multi-station velocity measurements for a subset of sources outside the Northern Hemisphere winter), but with a relatively sparse network of sources.
ASKAP could rapidly densify the IPS measurements at the sky location where an unusually strong ISEE detection was made, allowing the nature, location and morphology of the enhanced scattering to be determined.

Similarly, ASKAP can inform further ground-based observations as part of a Worldwide Interplanetary Scintillation Stations (WIPSS) Network \citep{2021cosp...43E2370B}.
For example, the location and motion of a CME as determined by ASKAP can facilitate the planning of IPS observations by International LOFAR \citep{2013A&A...556A...2V}, which has its own unique IPS observing capabilities \citep{2021cosp...43E1026F}.
Beyond IPS observations, ASKAP and other observations can assist in the scheduling and interpretation of radio observations (by ASKAP or by other instruments) designed to determine (via the Faraday Rotation (FR) and hence the Magnetic field orientation) the geoeffectiveness of CMEs.
IPS identifies exactly which points on the sky the CME is dense; FR measurements made at those  locations will have the highest signal-to-noise ratio.
We note that while LOFAR and the MWA cannot make IPS observations (at least in the weak scintillation regime) as close to the Sun as ASKAP, they can make FR measurements closer, where the FR is likely to be higher and therefore more detectable \citep{2012RaSc...47.0K08O}.

Finally, ASKAP IPS observations can feed directly into heliospheric modelling efforts which provide Space Weather Forecasts \citep[e.g.][]{1998JGR...10312049J, Jackson2020FrASS...7...76J}.
ASKAP measurements, being made relatively close to the Sun, and with a predicted density of sources much higher than any previous IPS observations, may be particularly useful for providing inner boundary conditions to simulations of the inner heliosphere \citep{2015SpWea..13..104J}.

\section{Acknowledgments}
The authors wish to acknowledge the support of A. Hotan, the broader ASKAP Operations team at CSIRO and the CRAFT team to make this test observation possible. 
The Australian Square Kilometre Array Pathfinder is part of the Australia Telescope National Facility which is managed by CSIRO. Operation of ASKAP is funded by the Australian Government with support from the National Collaborative Research Infrastructure Strategy. ASKAP uses the resources of the Pawsey Supercomputing Centre. Establishment of ASKAP, the Murchison Radio-astronomy Observatory and the Pawsey Supercomputing Centre are initiatives of the Australian Government, with support from the Government of Western Australia and the Science and Industry Endowment Fund. We acknowledge the Wajarri Yamatji as the traditional owners of the Murchison Radio-astronomy Observatory site. Ryan Shannon acknowledges support through Australian Research Council Future Fellowship FT190100155.

%% Bibliography
%% Author year style
\bibliographystyle{model5-names}
\biboptions{authoryear}
\bibliography{refs}

\end{document}